\documentclass[reprint,showpacs,amsmath,amssymb,showkeys,aps,prl]{revtex4-1}
\usepackage{natbib}
\usepackage{graphicx}
\usepackage{dcolumn}
\usepackage{bm}
\usepackage{hyperref}

\begin{document}
\title{Friction Boosted by Equilibrium Misalignment of Incommensurate Two-Dimensional Colloid Monolayers}
\author{Davide Mandelli$^1$, Andrea Vanossi$^{2,1}$, Nicola Manini$^{3,1,2}$, Erio Tosatti$^{1,2,4}$}
\affiliation{
$^1$ International School for Advanced Studies (SISSA),
     Via Bonomea 265, 34136 Trieste, Italy \\
$^2$ CNR-IOM Democritos National Simulation Center,
     Via Bonomea 265, 34136 Trieste, Italy \\
$^3$ Dipartimento di Fisica, Universit\`a degli Studi di Milano, 
     Via Celoria 16, 20133 Milano, Italy \\
$^4$ International Centre for Theoretical Physics (ICTP),
     Strada Costiera 11, 34014 Trieste, Italy
            }
\date{\today}
\begin{abstract}
Colloidal 2D monolayers sliding in an optical lattice are of recent importance as a frictional system.
In the general case when the monolayer and optical lattices are incommensurate, we predict two 
important novelties, one in the static equilibrium structure, the other in the frictional behavior under sliding.
Structurally, realistic simulations show that the colloid layer should possess in full equilibrium a small 
misalignment rotation angle relative to the optical lattice, an effect so far unnoticed but visible in 
some published experimental moir\'e patterns. Under sliding, this misalignment has the effect of boosting 
the colloid monolayer friction by a considerable factor over the hypothetical aligned case discussed so far. 
A frictional increase of similar origin must generally affect other incommensurate adsorbed monolayers and 
contacts, to be sought out case by case.
\end{abstract}
\maketitle
The mutual sliding of crystalline lattices offers, despite its apparently
academic nature, one of the basic platforms to understand the nanoscale and
mesoscale frictional and adhesion phenomena \cite{VANRMP}.
In one of the freshest developments, Bohlein and collaborators \cite{BOHL} showed that 
the sliding of a 2D crystalline monolayer of colloidal particles in an optical lattice 
provides unexpected information on elementary tribological processes in 
crystalline sliding systems with ideally controlled commensurabilities.
Given the scarcity of reliable and controllable frictional systems,
it is hard to overestimate the importance of such model systems
with full external control over all parameters including periodicity,
coupling strengths, and applied forces.
For this reason 2D monolayers in periodic lattices require a close theoretical study.
In this letter we describe two main surprises, one structural and one frictional,
 which emerge from realistic molecular dynamics simulations.
 We show first of all that incommensurate colloid islands naturally develop in full equilibrium a small 
 misalignment angle relative to the substrate. Second, sliding simulations demonstrate that the misaligned
 angular orientation increases significantly the dynamic friction with respect to the (hypothetical and unstable)
 aligned case.
 While both are important for colloidal monolayers, their potential impact extends in
 principle beyond the specific case, to a wider variety of systems where mutually incommensurate 2D lattices 
 are brought in static and then in sliding contact.
 
\begin{figure}[!t]
 \begin{center}
 \includegraphics[angle=0, width=0.46\textwidth]{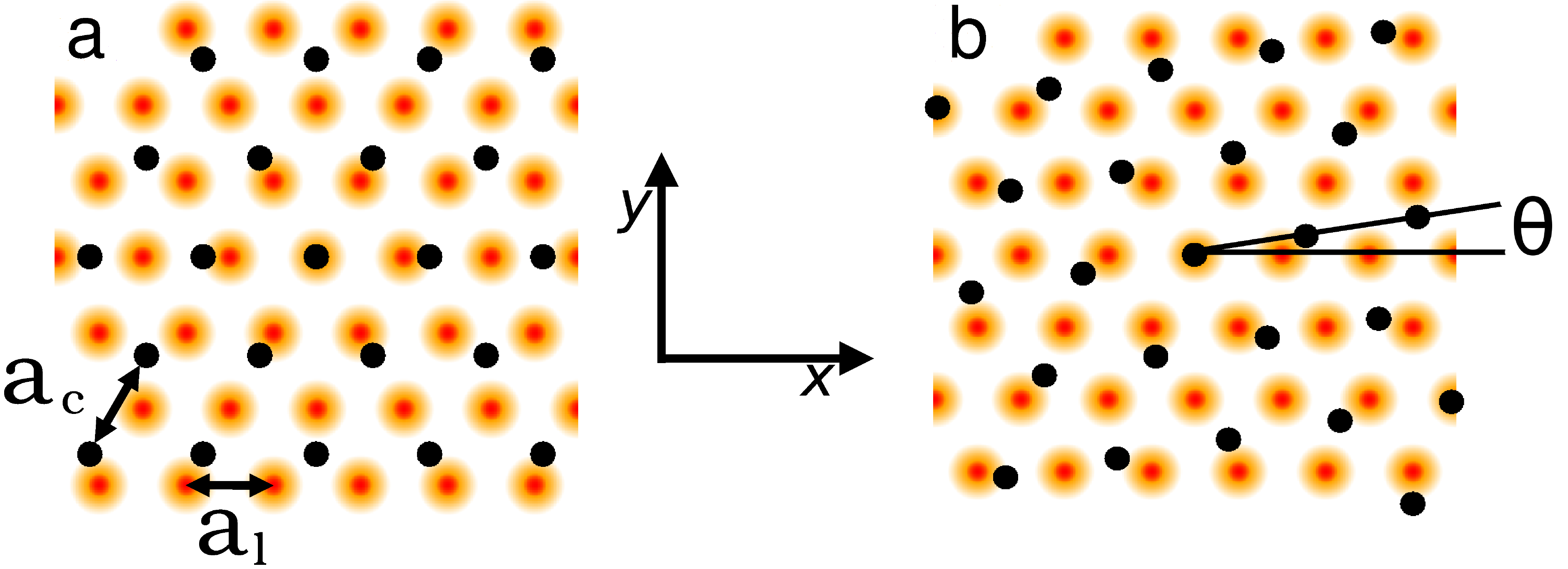}
 \caption{ \label{fig1}
 (Color online)
 (a) Schematic of a 2D colloidal lattice (black dots) aligned with an incommensurate 
triangular periodic potential mimicking the optical lattice (blurred spots represent potential minima).
 (b) A misaligned configuration rotated by an angle $\theta$. }
 \end{center}
 \end{figure}

In colloid monolayers, the 2D density may vary from ``underdense'' 
($\rho<1$, where $\rho=a_l/a_c$, with $a_l$ the spacing of the optical lattice, $a_c$ that of the colloid lattice, 
both triangular) to perfectly commensurate ($\rho=1$, one particle per potential well), to ``overdense'' ($\rho>1$), 
each with its specific sliding behavior. Both experiments \cite{BOHL} and theory \cite{MANI,DELLAGO} indicated 
that commensurate ($\rho=1$) friction is large, 
dropping to much lower values  for $\rho \neq 1$, where optical and colloid lattice are incommensurate.
This drop reflects the great mobility of the pre-existing misfit dislocations, also called kinks or ``solitons''.
In these studies the two lattices, colloid and optical, were silently assumed to be geometrically aligned, prior
to and during sliding. That assumption however is dangerous.
A long-known theoretical result suggests \cite{NOV}, for example, that a harmonic monolayer 
subject to an incommensurate periodic potential of weak amplitude $U_0$ may partly convert the  
misfit compressional stress to shear stress by an equilibrium geometric misalignment of the monolayer 
(see Fig.~\ref{fig1}) through a small rotation angle 
\begin{equation}
\label{eq:theta}
\theta_{\rm NM}=
\arccos\left(\frac{1+\rho^2(1+2\delta)}{\rho[2+\delta(1+\rho^2)]}\right),
\end{equation}
whose energy-lowering effect originates from a better interdigitation of the two lattices.
Independent of $U_0$, the rotation angle is nonzero in this approximation provided the transverse 2D 
sound velocity $c_T$ is sufficiently smaller than the longitudinal $c_L$, 
precisely if $\delta=(c_L/c_T)^2-1>\rho^{-1}$.
While this kind of rotated epitaxy has been addressed
experimentally \cite{SHA,ARU} and theoretically \cite{TOM} for adsorbed rare-gas monolayers, its 
possible presence in colloidal monolayers was so far unsuspected. More importantly in the context of 
sliding friction, the tribological impact of an {\it  equilibrium} geometrical misalignment is unexplored 
in any incommensurate system. The  externally forced rotation turning a {\it commensurate} layer into 
incommensurate is well known to reduce friction, as exemplified by the sliding of graphene flakes on graphite \cite{FRE,FIL}.
Different as these two cases are, a possible expectation might
be that the equilibrium geometry, alignment of commensurate layers, 
or misalignment of incommensurate layers,
should always exhibit a higher friction relative to forcedly
rotated ones, since the optimal $T=0$ geometry must in
every case corresponds to a closer interdigitation of the
two lattices.

First, let us consider structural alignement. Using the same methods as in Ref. \onlinecite{MANI}, we model 
the colloidal system in an optical lattice as a 2D monolayer of $N_p$ point-like classical particles,
mutually repelling through a screened Coulomb potential $U_{\rm pp}(r_{ij})$ while immersed in a
static 2D triangular lattice potential $W({\bf r}_i)=U_0w({\bf r}_i)$ where $w({\bf r})$ is a dimensionless
periodic function of spacing $a_{l}$ ( as specified in Supplemental Material ({\it SM}) ), and $U_0$ is the
amplitude (``corrugation'') parameter. 
The Hamiltonian is thus 
\begin{equation}
  H = \sum_{i=1}^{N_p}  \left[ U_0w({\bf r}_i) +
    \frac 12  
    \sum_{j \neq i} U_{\rm pp}(r_{ij}) \right].
\end{equation}
The particles are confined to the $(x,y)$ plane, and subjected to either planar periodic boundary 
conditions (PBC) in a given area $A$, or alternatively to an additional
Gaussian confining potential $G(r)=-A_{G}\exp(-r^2/\sigma_G^2)$,
in which case they form an island with open boundary conditions (OBC).
Temperature is generally set to zero, because the results are clearer
and require less statistics in this limit. Finite temperature results are otherwise
not essentially different, as shown along with details of
optimization protocols in Supplemental Material ({\it SM}). 
Within the confinement region the 2D particles crystallize in a triangular
lattice of mean spacing $a_c$, with modulations induced by the periodic
potential (in the OBC case there is also a confinement-induced
variation between a dense center and a sparser periphery). 
Here and in the rest we shall focus for specificity 
on the underdense incommensurate case $\rho=3/(1+\sqrt{5})\simeq 0.927$;
the physics with different values of $\rho \neq 1 $ including $\rho>1$ is 
qualitatively similar, as briefly discussed in {\it SM}. 

\begin{figure}[!t]
 \begin{center}
 \includegraphics[angle=0, width=0.46\textwidth]{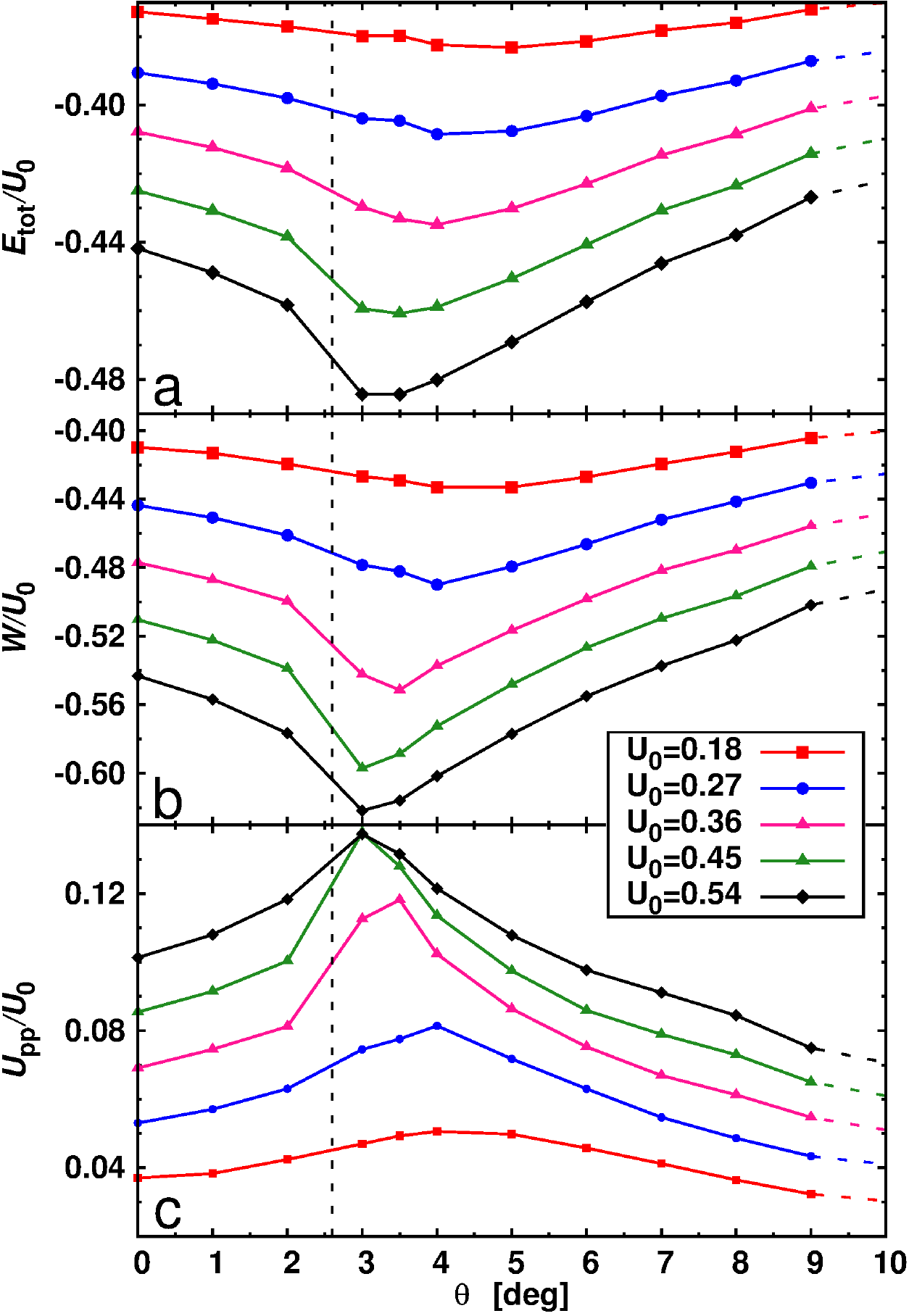}
 \caption{ \label{fig2}
   (Color online)
   Relative static energy of structure-optimized colloid islands (OBC) as a function of the trial
   rotation angle $\theta$.
   (a) Total energy per particle $E_{\rm tot}$.
   (b) Periodic-potential (corrugation) contribution
   $W$ to $E_{\rm tot}$.
   (c) Interparticle interaction contribution $U_{\rm pp}$ to $E_{\rm tot}$.
   Curves correspond to increasing corrugation amplitude $U_0=0.18-0.54$.
   Energies are measured relative to that of the colloidal
   monolayer at rest and at $U_0=0$. Dashed line: ideal NM angle,
   Eq.~\eqref{eq:theta}, $\theta_{\rm NM}\approx2.6^\circ$. }
 \end{center}
 \end{figure}

We determined, by means of careful energy minimization, the optimal $T=0$ 2D geometry of 
all $N_p$ particles for increasing corrugation strength; shown here are the 
results for $U_0=0.1-0.6$, where the effects are particularly clear.
The final, optimal geometry of the 2D colloid lattice initially aligned at $\theta=0^{\circ}$ 
with the optical lattice axes is found to be misaligned, with a small rotation angle 
$\theta_{\rm opt}\simeq 2.3^\circ$ in PBC calculations.
This rotation realizes a better interdigitation with the optical lattice, and occurs
spontaneously during the simulation at the cost of creating the dislocations required by the PBC 
constraints (see {\it SM}). 
More detailed energy minimizations were done in OBC, which do not have the same problem.
Since the 30,000 particle islands are too large to spontaneously rotate,
we carried out simulations starting from a prearranged sequence of misalignment angles.
The total energy minimum versus $\theta$ confirms, as 
shown by Fig.~\ref{fig2}, the structural misalignment angle at equilibrium, with a magnitude in the 
range $3^{\circ} < \theta_{\rm opt} <5^{\circ}$ (depending on the optical lattice strength $U_0$)
generally somewhat larger than in PBC, where angular constraints hinder the misalignment.

These structural results compare instructively with those expected from the
harmonic model \cite{NOV}.
In our case a 2D phonon calculation for the monolayer yields a sound velocity ratio $c_L/c_T=1.806$, 
larger than the theoretical threshold value $(1+\rho^{-1})^{1/2}\simeq 1.442$.
The corresponding theoretical misalignment $\theta_{\rm NM}\simeq 2.6^\circ$ is in qualitative agreement with the 
more realistic simulation result.
Figure~\ref{fig2}b,c shows how the two pieces which compose the total energy, namely the periodic lattice
energy part $W=\langle W({\bf r}_i) \rangle$ controlled by the corrugation amplitude $U_0$ and the interparticle 
interaction $U_{\rm pp}=\langle U_{\rm pp}(r_{ij})\rangle$, behave. Misalignment raises the interparticle energy, 
but that cost is overcompensated, by a factor 2, by a corrugation energy gain. The incommensurate colloid 
static equilibrium structure is misaligned because that permits a better interdigitation with the 
optical lattice.

 \begin{figure}[!t]
 \begin{center}
 \includegraphics[angle=0, width=0.46\textwidth]{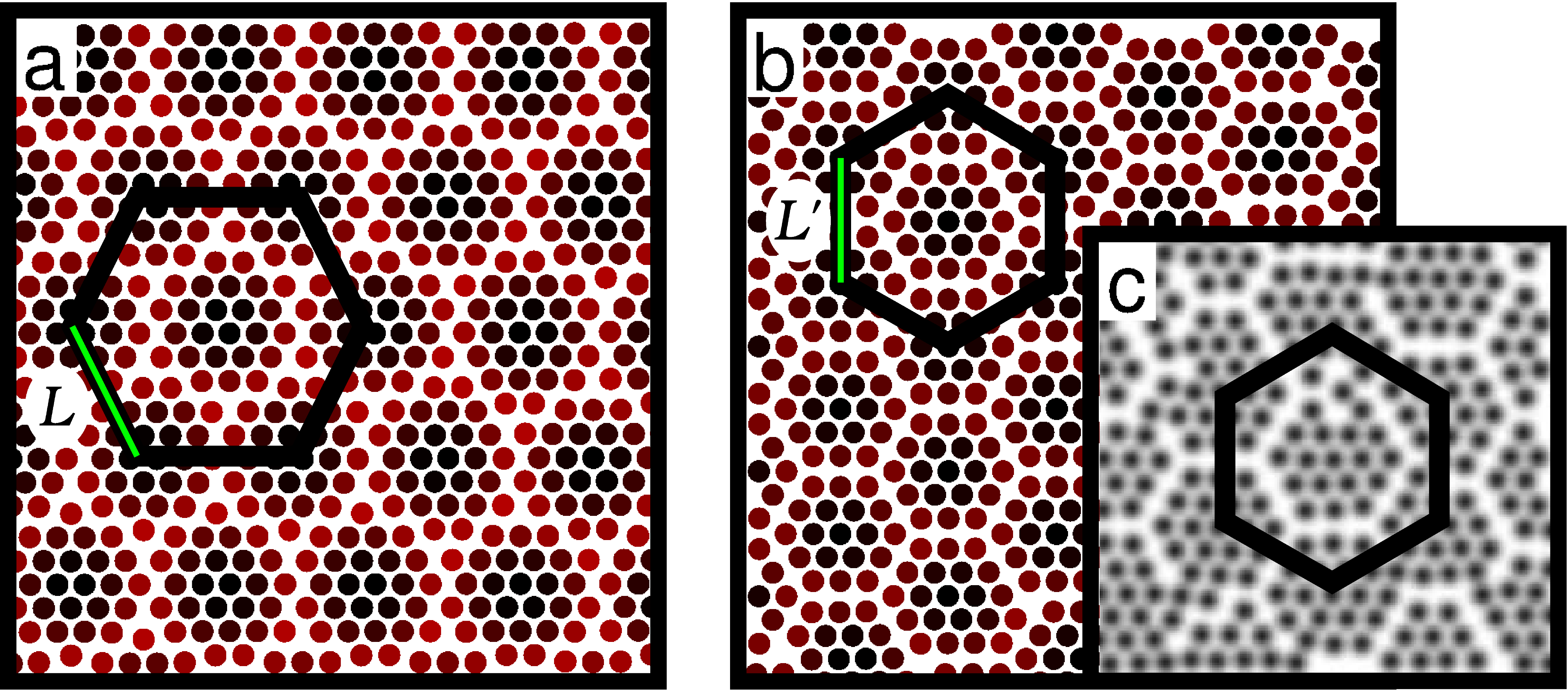}
 \caption{\label{fig3}
   (Color online)
   Equilibrium configurations obtained for $\rho\simeq 0.83$ and $U_0=6.3$.
   (a) Unrotated  $\theta=0^{\circ}$; (b) optimally rotated
   $\theta_{\rm opt} =7^{\circ}$.
   Dark/light colloids enjoy best/worse $W$.
   Only the central part of the island, optimized in OBC, is shown.
   (c) Experimental geometry for the same $\rho$, adapted from
   Ref.~\onlinecite{EPL}, where both the moir\'e angles and spacing compare
   directly with (b) rather than (a). }
 \end{center}
 \end{figure}

Even if our predicted misalignments are small, their experimental existence is easily revealed, 
because the rotation angle $\theta$ between two lattices is highly amplified
by the moir\'e pattern, which rotates relative to the periodic potential
lattice by an angle $\psi$ satisfying the geometric relation
 $\cos{\theta} = \rho^{-1}\sin^2{\psi} + \cos{\psi}
\, \left[1-\rho^{-2} \sin^2{\psi}\right]^{1/2}$ \cite{BOHR}.
As the moir\'e superlattice rotates by $\psi$, its spacing $L$ also
decreases \cite{GUINEA} from its aligned value of about $L$=$a_c$$\rho$/(1-$\rho$) 
to its rotated value of $L'$=$a_c/\sqrt{1+\rho^{-2}-2\rho^{-1}\cos\theta}$.
As an example, Fig.~\ref{fig3} reports the structures of the artificially unrotated and of the optimally
rotated ($\theta_{\rm opt}=7^\circ$) configurations calculated for $\rho\simeq 0.83$ and $U_0=6.3$ 
(parameters believed to be appropriate to experiments in Ref.~\onlinecite{EPL}) in comparison with one another and 
with the corresponding experimental structural moir\'e.
Both the orientation and spacing of Fig.~\ref{fig3}c agree with the
$\theta_{\rm opt}=7^\circ$ but not with the $\theta_{\rm opt}=0^\circ$ pattern, 
proving that the misalignment was actually present in that experiment.

\begin{figure}[t]
 \begin{center}
 \includegraphics[angle=0, width=0.46\textwidth]{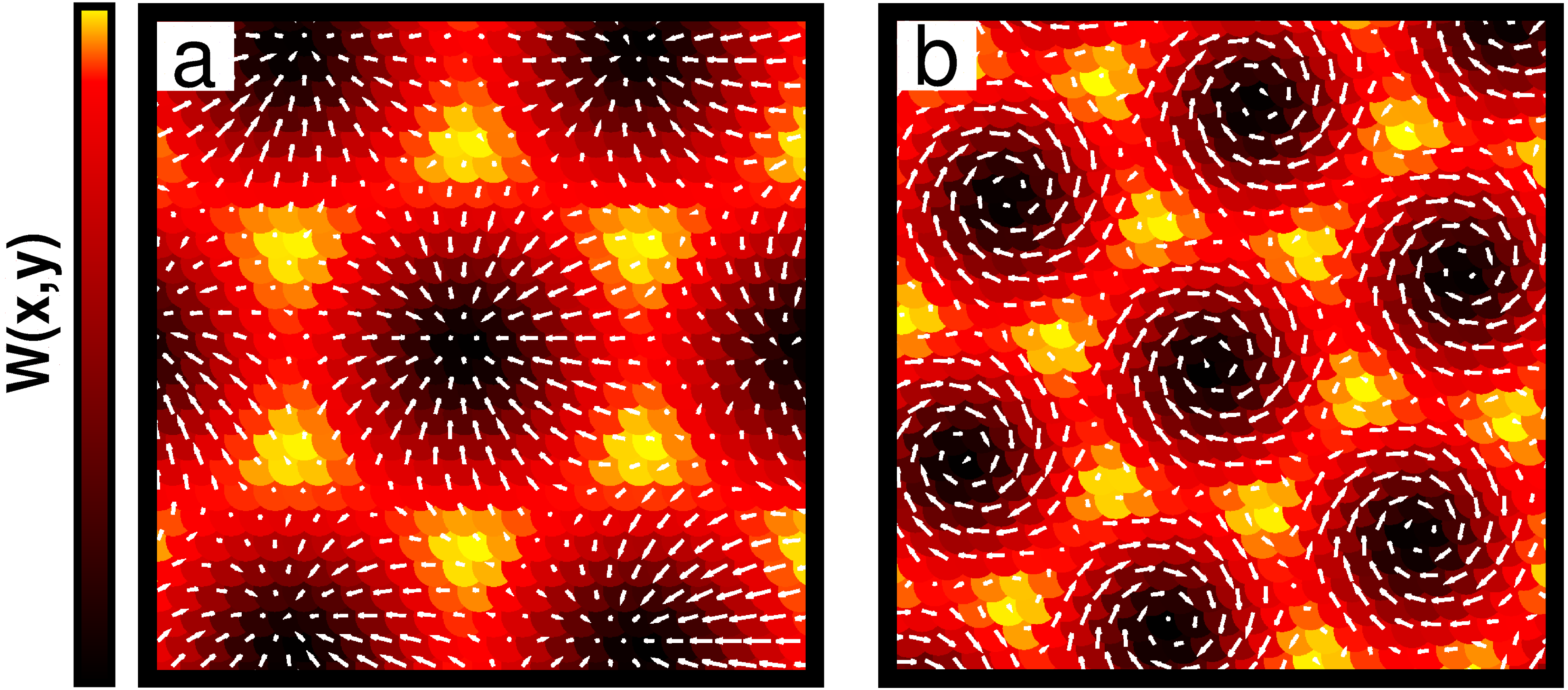}
 \caption{ \label{fig4}
   (Color online)
   Moir\'e patterns of the particle monolayer's central region as obtained
   in OBC for $\rho=0.927$, $U_0=0.27$,
   (a) for $\theta=0^{\circ}$ (moir\'e angle $\psi=0^{\circ}$)  and
   (b) for the optimal $\theta=4^{\circ}$
   (moir\'e angle $\psi\approx 15^{\circ}$).
   Each dot indicates a particle in the unrelaxed configuration,
   colored according to the local corrugation potential
   $W({\bf r})$: dark for potential minima, bright for maxima.
   White arrows show the displacements of each particle from the ideal
   triangular lattice to the fully relaxed configuration, magnified 15 times.
   The compression-dilations at $\theta=0^{\circ}$ are turned into largely
   shear, vortex-like displacements at $\theta_{\rm opt}=4^{\circ}$. }
 \end{center}
 \end{figure}

The particle static displacements associated with the optical lattice potential are also enlightening.
Figure~\ref{fig4} shows the moir\'e pattern of a small portion of the monolayer island ($\rho=0.927$, $U_0=0.27$)
for $\theta=0^{\circ}$ and for $\theta_{\rm opt}=4^{\circ}$,
corresponding to a moir\'e angle $\psi\approx 15^{\circ}$.
Particle displacements, designated by arrows, change from radial longitudinal compression-dilations 
to mixed shear-longitudinal, vortex-like displacements upon optimal misalignment.
A large 2D bulk modulus and a weak shear rigidity of the crystalline monolayer are crucial factors increasing
the extent of the shear distortions, therefore enhancing the lattice misalignment.

%
%

We come to our second point, i.e., the forced sliding 
of the particle monolayer over the periodic corrugation and the
associated frictional losses.   
The shear distortions and the corresponding increased interdigitation at the
optimal misalignment angle $\theta_{\rm opt}$ are expected to affect the
sliding of the particle lattice over the periodic potential.
Sliding is realized by a flow of the soliton superstructure, 
accompanied by dissipation as part of the work goes into
soliton-related time-dependent distortions of the 2D lattice.
That work will change once the nature (longitudinal to shear), 
orientation ($0^{\circ}$ to $\psi$), periodicity ($L$ to $L'$) change with 
$\theta$ ($0^{\circ}$ to $\theta_{\rm opt}$).
We determine the magnitude of the expected friction change by simulating the overdamped 
sliding dynamics of the OBC island over a range of $\theta$ values, so as to assess the frictional
effect of misalignment near its optimal value.
We applied an external driving force $F_d$ acting on each particle, 
slowly varying to-and-fro as a function of time, mimicking the experimental drag force 
$\eta v_d$ induced by a fluid of viscosity $\eta$ and slowly back and forth time-dependent 
speed $v_d$ \cite{BOHL} (details in {\it SM}). Despite a nonzero torque, generally present for 
all preset angles that differ from $\theta_{\rm opt}$ (and from zero) the misalignment angle 
did not have the time to change appreciably in the course of the simulation. 
Under sliding, the frictional power dissipated per particle was calculated as \cite{MANI}
\begin{equation}
\label{eq:power}
p_{\rm fric}= {\bf F_d} \cdot \langle {\bf v}_{\rm cm}\rangle-\eta |\langle {\bf v}_{\rm cm}\rangle|^2 = (\eta/N_p) \Sigma_i \langle| {\bf v}_{\rm i} - {\bf v}_{\rm cm}|^2\rangle
\end{equation}
where ${\bf v}_{\rm cm}$ is the center-of-mass velocity, ${\bf v}_{\rm i}$ is the velocity 
of particle $i$, and brackets denote steady-state averages. 
Due to the confining envelope potential, the lattice spacing $a_{c}$, close to constant 
in the central part, increases toward the periphery, where colloids also tend to be pinned by the corrugation.  
To address properties of mobile colloids at a well-defined density, trajectories were
analyzed, as also done in experiments, considering only particles belonging to the central portion of the
island, in our case a square of size $80\times80\,a_c^2$. 
The duration of the sliding simulations was fixed by requiring a total
center of mass displacement not smaller than $\Delta x_{\rm  cm}\approx 2-3\,a_{c}$.
 \begin{figure}[!t]
 \begin{center}
 \includegraphics[angle=0, width=0.46\textwidth]{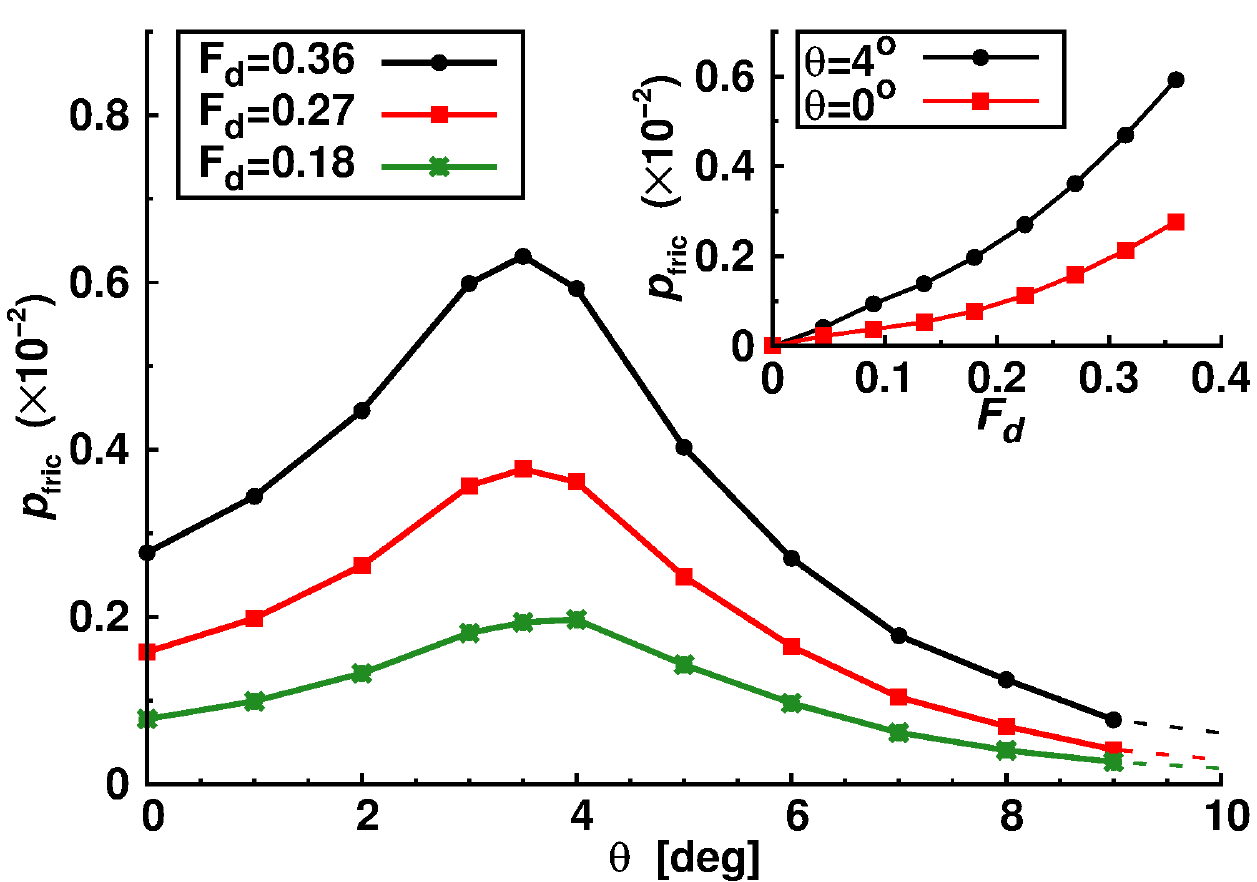}
 \caption{ \label{fig5}
   (Color online)
   Dissipated friction power $p_{\rm fric}$ as a function of the trial misalignment
   angle $\theta$ ($U_0=0.27$).
   Three curves are reported corresponding to increasing values of the
   driving force $F_d=0.18,\ 0.27,\ 0.36$.
   The inset shows $p_{\rm fric}$ as a function of $F_d$ for two values
   of $\theta$. }
 \end{center}
 \end{figure}

\begin{figure}[!t]
 \begin{center}
 \includegraphics[angle=0, width=0.46\textwidth]{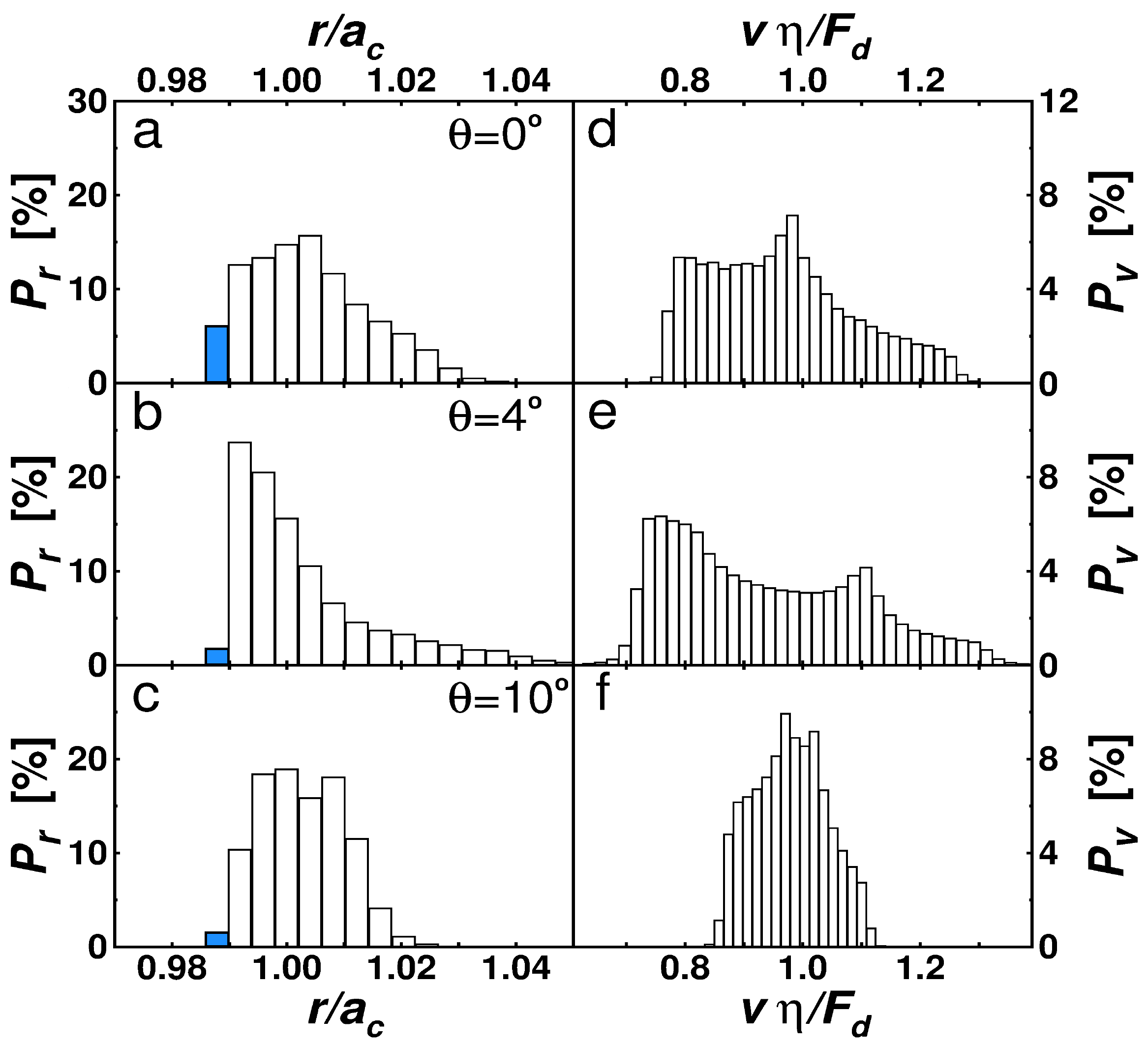}
 \caption{\label{fig6}
   (Color online)
   (a-c) The distribution $P_r$ of nearest-neighbor distances in the
   static relaxed configurations ($F_d=0$, $U_0=0.27$) at
   rotation angles $\theta=0^{\circ}$, $\theta=\theta_{\rm opt}=4^{\circ}$, 
   and $\theta=10^{\circ}$.
   (d-f) Corresponding velocity distribution $P_v$ of particles sliding
   under an external force $F_d=0.36$.}
 \end{center}
 \end{figure}

Figure~\ref{fig5}, our main dynamical result, shows that friction is increased by a very
substantial factor by misalignment relative to alignement, reaching a maximum of about two 
at the optimal angle $\theta_{\rm opt}$, subsequently dropping for
larger angles where the energy gain and static distortion magnitude also drop.
The physical reason for the frictional peak at $\theta_{\rm opt}$ 
can be further appreciated by looking at the particles' steady state velocity distribution
$P_v$ and at the corresponding static interparticle spacing distribution
(at zero velocity) $P_r$, both shown in Fig.~\ref{fig6} for increasing $\theta$.
The important points here are that small interparticle distances are energetically costly, 
and that a large spread of velocities relative to the center-of-mass denotes larger frictional dissipation, according to the RHS of Eq.~\eqref{eq:power}.
At perfect alignment, short distances (colored column) are very frequent, which is energetically costly. At the same time the spread of velocities is moderate and so is friction.
In the optimally misaligned case $\theta_{\rm opt}$ instead, the shortest distance becomes 
less frequent, thus reducing energy as we already know.
At the same time however $P_v$ develops longer tails at lower and higher particle velocities, both of which increase friction.
At $\theta > \theta_{\rm opt}$ finally the velocity spread drops and 
so does friction, the monolayer sliding less and less affected by corrugation. 
%
%


In conclusion, colloid monolayers in an incommensurate optical lattice develop, in full equilibrium 
and with realistic parameters, a small-angle structural misalignment, quite evident in moir\'e patterns 
such as those of Fig.~\ref{fig3}.
Upon forced sliding, this misalignment can considerably increase the sliding friction,
directly extractable from the colloid drift velocity in experiment, using Eq.~\eqref{eq:power}, over the hypothetical aligned geometry.

The present results and understanding naturally extrapolate to the sliding
of misaligned incommensurate lattices in contact such as, for example,
physisorbed rare-gas or molecular submonolayer islands \cite{SHA,ARU,TOM}.
An interesting side aspect is in this case that misalignment transforms the orientation angle of a physisorbed 
island, generally assumed to be fixed, into a continuous and possibly dynamical variable. The inertial sliding 
friction of such islands determines the inverse slip time in
quartz crystal microbalance experiments \cite{KRI}, 
whose data must, at least in some cases, embed the frictional enhancement caused by misalignment when 
present.
Even though the time needed to diffuse-rotate a $\sim 100$~nm-size island may be
exceedingly large, the orientation angle distribution of islands will
usually, under either stationary or sliding conditions, be continuous rather
than delta-function like.
The general unavailability of 2D lattice orientation of incommensurate 
rare gas islands (as opposed to full monolayers, whose epitaxy generally 
differs) poses at present an obstacle to the investigation of these effects, 
which must nonetheless be considered as generically present and
effective. The possible local misalignment of incommensurate 3D 
crystals in contact and its potential role in sliding friction 
is an even less explored, but interesting issue which remains 
open for future consideration.

This work was mainly supported under the ERC Advanced Grant
No.\ 320796-MODPHYSFRICT,
and partly by the Swiss 
National Science Foundation through a SINERGIA contract CRSII2\_136287,
by PRIN/COFIN Contract 2010LLKJBX 004, 
and
by COST Action MP1303.


\end{document}


\preprint{APS/123-QED}
\title{Friction Boosted by Equilibrium Misalignment of Incommensurate Two-Dimensional Colloid Monolayers-- Supplemental Material}
\author{Davide Mandelli$^1$, Andrea Vanossi$^{2,1}$, Nicola Manini$^{3,1,2}$, Erio Tosatti$^{1,2,4}$}
\affiliation{
$^1$ International School for Advanced Studies (SISSA),
     Via Bonomea 265, 34136 Trieste, Italy \\
$^2$ CNR-IOM Democritos National Simulation Center,
     Via Bonomea 265, 34136 Trieste, Italy \\
$^3$ Dipartimento di Fisica, Universit\`a degli Studi di Milano,
     Via Celoria 16, 20133 Milano, Italy \\
$^4$ International Centre for Theoretical Physics (ICTP),
     Strada Costiera 11, 34014 Trieste, Italy
            }
\date{\today}
\maketitle
\section{Model details}
We describe the colloidal particles as classical point-like objects interacting 
via a repulsive Yukawa potential
\begin{equation}
\label{eq:Yuk}
 U_{\rm pp}(r)=\frac{Q}{r}\exp\left(-r/\lambda_{\rm D}\right),
\end{equation}
where $r$ is the interparticle distance, $Q$ is the coupling strength, and
$\lambda_{\rm D}$ is the Debye screening length.
We restrict particle motion to two dimensions, where colloids at rest form
a triangular lattice of spacing $a_c$ in the $(x,y)$ plane.
The periodic triangular corrugation potential
\begin{eqnarray}
\label{eq:tripot}
 W({\bf r})&=&-\frac{2}{9}U_0\left[\frac{3}{2}+2\cos\frac{2\pi x}{a_l}\cos\frac{2\pi y}{\sqrt{3}a_l}
            +\cos\frac{4\pi y}{\sqrt{3}a_l}\right] \\ \nonumber
           &=& U_0w({\bf r})
\end{eqnarray}
has strength $U_0$ and periodicity $a_l$, mimicking the experimental laser-induced optical lattice.
The ratio $\rho$=$a_l/a_c$ defines their relative commensurability. Here we focus on mismatched cases only,
namely underdense incommensurate UI ($\rho<1$) and overdense incommensurate OI ($\rho>1$).
A detailed discussion of the approximations behind this model can be found
in Ref. \onlinecite{MANI}.

Simulations are carried out using either constant-area periodic boundary
conditions (PBC), or open boundary conditions (OBC) with a confining potential.
In PBC, the purely repulsive particles are naturally confined and the total
potential energy of $N_p$ particles is
\begin{equation}
\label{eq:Hpbc}
H= \sum_{i=1}^{N_p}  \left[ W({\bf r}_i) + \frac 12 \sum_{j \neq i} U_{\rm pp}(r_{ij}) \right].
\end{equation}
We use a triangular-lattice supercell defined by primitive vectors ${\bf a}_1$=$L(1,0)$, 
${\bf a}_2$=$L(0.5,\sqrt{3}/2)$ initially filled with a
regularly spaced triangular monolayer of lattice constant
$a_c=1$, taken as our length unit. The supercell edge $L$ and the corrugation
potential wavelength are then chosen to achieve a given overall
periodicity of the colloidal lattice and the periodic potential. After a
number of trials, we settled on simulations using $L=89\,a_c$ ($N_p=7,891$)
and $\rho=89/96\approx0.927$, corresponding to the fifth approximant in the
continuous fraction expansion of $3/(1+\sqrt{5})$.
 \begin{figure}
 \begin{center}
 \includegraphics[angle=0,width=0.45\textwidth]{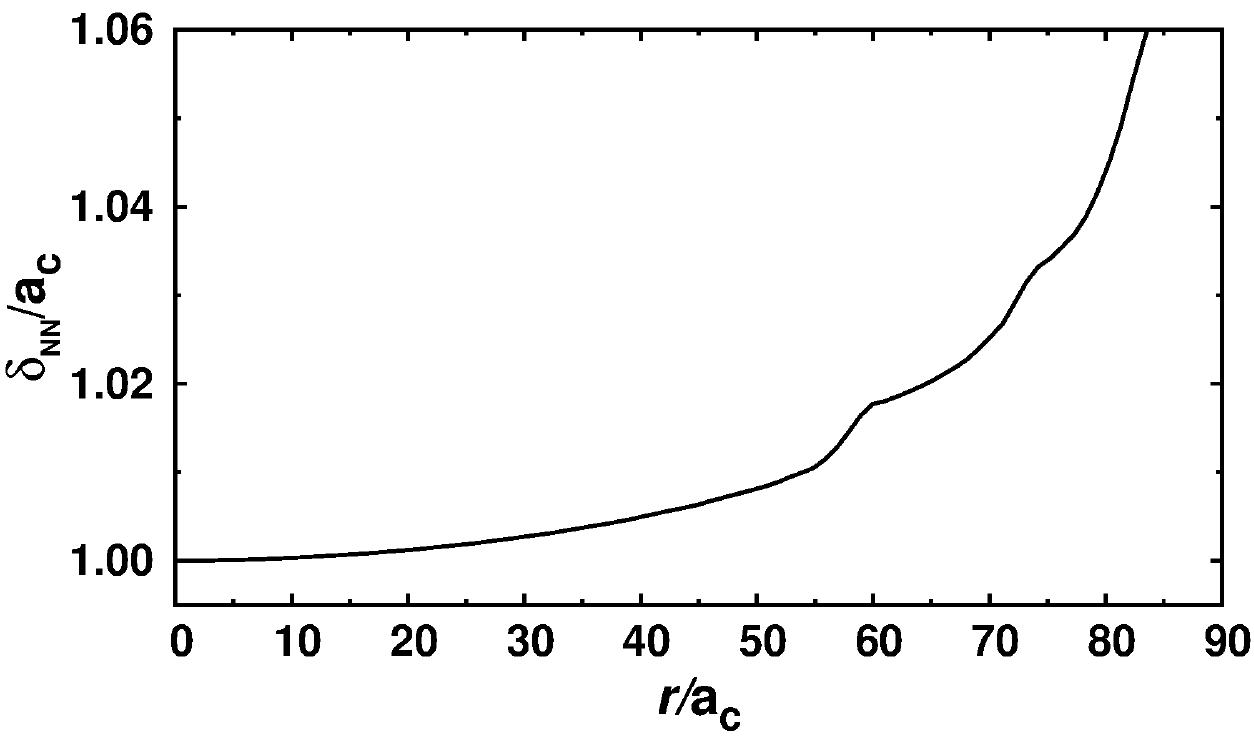}
 \caption{\label{fig1}
          Average nearest neighbor spacing as a function of the distance
          from the center of the model colloidal island containing $N_p=28,837$
          particles, optimized at rest in OBC at $T=0$, $U_0=0$, and $\rho\simeq 0.927$.
          The kinks at large radii are connected with the 
          presence of line defects appearing in the relaxed configuration
          as a result of the growing inhomogeneity.
         }
 \end{center}
 \end{figure}

In OBC a large square supercell of size $L=500\,a_c$ is partly filled by
a circular island of $N_p\simeq 30,000$ particles, moving in the total external potential
\begin{equation}
\label{eq:Vext}
V_{\rm ext}({\bf r})=G({\bf r})\left[-A_G+W({\bf r})\right],
\end{equation}
where
\begin{equation}
\label{eq:Gaus}
G({\bf r})=\exp\left(-\frac{|{\bf r}|^2}{2\sigma_G^2}\right),
\end{equation}
is an unnormalized confining Gaussian of large width $\sigma_{G}$.
As suggested by experiment \cite{BOHL} both the confining amplitude $A_G$
and the corrugation potential $W({\bf r})$ are controlled by the same Gaussian modulation.
By fixing $Q=10^{13}$, $\lambda_{\rm D}=0.03$, $A_{G}=1200$,
$\sigma_{G}=1200$, the Gaussian confinement and the interparticle repulsion
have been balanced so as to yield an interparticle distance $a_c\simeq
0.983$ at the island center.
As shown in Fig.~\ref{fig1}, the lattice spacing is fairly constant in the
central region and increases by a few percent toward the island periphery.
The total potential energy has the same expression as Eq.~\eqref{eq:Hpbc}, 
but with $V_{ext}({\bf r}_i)$ in place of $W({\bf r}_i)$.
We carry out simulations at different lattice mismatch
$\rho=0.927,\ 1.08,\ 0.84$ set by choosing appropriate values of $a_l$
given the central colloidal crystal spacing $a_c$.

The equation of motion for the $j$-th particle displacement ${\bf r}_j$ is
\begin{equation}
\label{eq:dyn}
 m\ddot{{\bf r}}_j+\eta(\dot{{\bf r}}_j-{\bf v}_d)=
-\nabla_{{\bf r}_j}\left[ \sum_{i\neq j}U_{\rm pp}(r_{ij})+V_{\rm ext}({\bf r}_j) \right],
\end{equation}
where ${\bf v}_d$ is the drift velocity, giving rise to a Stokes' drag force ${\bf F}_d$=$\eta{\bf v}_d$, 
experienced by each particle \cite{MANI}.
The overdamped dynamics typical of this system is achieved by adopting a
sufficiently large value of the damping coefficient $\eta$.
We use $\eta=28$ in all simulations.
All results are expressed in terms of the basic units defined in Table~\ref{tab1}, where values 
are inspired by those in the experiment by Bohlein \cite{BOHL}.
Figure 3 in the paper refers however to a different experimental setup \cite{EPL}
corresponding to a softer colloidal crystal at a larger lattice spacing $a_c\simeq6.5~\mu$m.
In that case a more appropriate value for the basic energy unit is $E\simeq k_BT_{\rm room}=4.1 \times10 ^{-21}$~J.
 \begin{center}
 \begin{table}[!t]
 \renewcommand{\arraystretch}{1.3}
 \begin{tabular}{l l}
 \hline
 Model Expression &Typical Value \\
 \hline
 Viscous coefficient $\eta$                                           &   $6.3\times10^{-8}$~kg/s \\
 Length $a_c$                                                         & $5.7~{\rm\mu}$m             \\
 Force  $F=Q e^{-a_c/\lambda_{\rm D}}/(\lambda_{\rm D} a_c)$ \quad  &  $20$~fN                  \\
 Energy $E=Fa_c$                                                      &   $1.1\times10^{-19}$~J   \\
 Velocity $V$=$F/\eta$                                                &   $0.3~\mu$m/s           \\
 Power $P$=$F^2/\eta$                                                 &   $6.3\times10^{-21}$~W   \\
 Time $t_0=\eta a_c/F$                                                &  $18$~s                   \\
 Mass $m=\eta^2a_c/F$                                                 &   $1.1\times 10^{-6}$~kg   \\
 \hline
 \end{tabular}
 \caption{\label{tab1}
         Basic units for various quantities in our model and typical values
         mimicking the setup of Ref. \onlinecite{BOHL}.
         }
 \end{table}
 \end{center}

\section{Simulation protocols -- statics}
\begin{figure}[!t]
 \begin{center}
 \includegraphics[angle=0,width=0.45\textwidth]{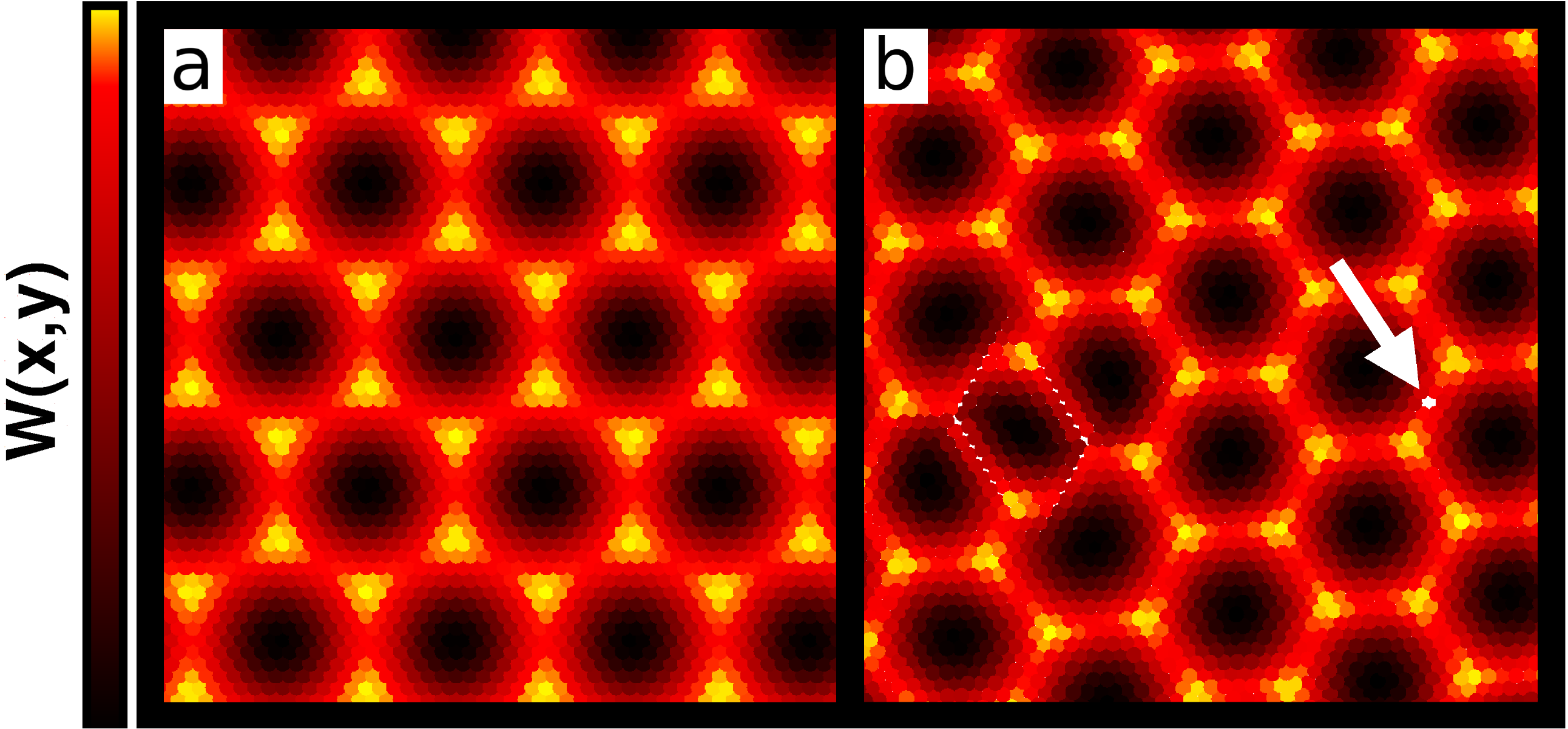}
 \caption{\label{fig2}
   A small region of the particle system optimized at rest in PBC.
   Particles are colored according to the local substrate potential
   $W(x,y)$.
   (a) Aligned ($\theta=0^\circ$) configuration obtained at $U_0=0.09$.
   (b) Configuration obtained at $U_0=0.18$, showing a spontaneous
   misalignment angle $\theta\simeq 2.3^\circ$, whose effect is greatly
   magnified in the corresponding moir\'e pattern, rotated by
   $\psi\approx30^\circ$ with respect to (a).
   An extended defect (at the left side) and a vacancy (pointed at by the
   arrow) appear, adapting the misaligned particle lattice to the
   constraint introduced by the PBC. }
 \end{center}
\end{figure}

In PBC the equilibrium configuration at each individual value of $U_0$ is generated as follows.
We start with a perfect colloidal crystal at $\theta=0^\circ$, as sketched in Fig.~1a of the paper.
A first series of zero-temperature ($T=0$) damped molecular-dynamics
simulations is carried out increasing the substrate potential amplitude in
steps $\Delta U=0.01$ from $U_0=0\rightarrow 0.5$, relaxing the monolayer
at each step until no appreciable movement of the colloids is observed.
A second series of configurations is generated starting with a corrugation
amplitude $U_0=0.5$, decreasing it in steps $\Delta U=-0.05$ down to
$U_0=0$ and performing an annealing protocol from $T=1\rightarrow0$ at each step.
The stability of each of these annealed structures is subsequently tested
by further carrying out parallel $T=0$ relaxations varying the corrugation
amplitude in steps $\Delta U= \pm 0.01$ (both upward and downward), until
$U_0=0$ or $U_0=0.5$ are reached.
For each value of $U_0$, among all the configurations generated from these
two procedures we eventually select the one with the lowest total energy.
The resulting lowest-energy configuration is usually aligned for small
corrugation $U_0$, and develops spontaneous misalignment and defects
for large enough $U_0$, see Fig.~\ref{fig2}. 

In OBC, equilibrium configurations are generated starting from a circular
island cut out of a perfect triangular crystal ($a_c=1$) and fully relaxed
inside the confining potential ($U_0=0$). The density profile resulting from 
the $U_0=0$ relaxation is drawn in Fig.~\ref{fig1}.
We then carry out damped-dynamics simulations at $T=0$ increasing the
corrugation $U_0$ in steps $\Delta U=0.02$.
We follow the same protocol starting from islands rotated at different
initial misalignment angles $\theta_i$, ranging between $0^\circ\leq\theta_i\leq 30^\circ$.
From this procedure, we obtain energy curves like those in Fig.~2 of the
paper, whose minima determine the optimal static configurations and
equilibrium misalignment angle $\theta_{\rm opt}$.
Note that in this process the final orientation $\theta_f$ of the island
at the end of optimization is not identical to the initial angle $\theta_i$; however, it
usually remains very close. Results are then given as a function of $\theta=\theta_i$.
All relaxed configurations with $\theta\neq\theta_{\rm opt}$ represent
locally stable (and generally metastable) states of the system.
\section{Simulation protocols -- sliding dynamics}
In real sliding experiments the driving force is the viscous drag exerted
on each colloid by a slow fixed-amplitude oscillatory motion of the experimental cell.
The slow time scale yields long time intervals with constant drag force. 
Our dynamical simulations are therefore carried out by applying to each
particle the same external force $F_d=\eta v_d$, usually along the $x$ axis.
$F_d$ is kept constant during a finite time $t_F$ inversely proportional to
the force value $F_d$ itself (and thus the overall sliding speed).
After this time, the force sign is reversed for the same time duration.
The particle motion being overdamped, there are no effects of inertia.
The product $F_d\,t_F$ is chosen so that an isolated unconfined particle
would move by a few lattice spacings $F_d\,t_F/\eta\approx 2-3\,a_c$ typically.

When we study the sliding of the monolayer in OBC, the Gaussian confining
potential is kept fixed in time.
Only the homogeneous central part of the island is sliding,
whereas the boundary tends to remain pinned during the slow oscillatory cycle.
In order to exclude undesired edge effects, we select a square central region 
of size $80\times80\,a_c^2$, containing $\sim 7,500$ particles.
Aiming at describing steady-state sliding, we discard an initial transient
of approximately 30\% of the simulation time and compute averages
considering particles which at $t=0$ were residing in this central block.

Following Ref. \onlinecite{MANI}, the dissipated friction power is
calculated according to Eq.~(3) in the paper.
Curves reported in Fig.~5 of the paper are obtained by applying this
procedure to each of the misaligned static configurations produced
following the protocol described earlier.
We verified that due to the relatively large size of the island and the
short duration of the simulations the initial angular orientation of the
island is preserved during the sliding trajectory (even at finite temperature).
We carried out mainly $T=0$ simulations integrating the equations of motions 
using an adaptive Runge-Kutta algorithm.
When needed, thermal (Brownian) motion is simulated by adding suitable
random Gaussian forces in a Langevin approach.
\begin{figure}[!t]
 \begin{center}
 \includegraphics[angle=0,width=0.45\textwidth]{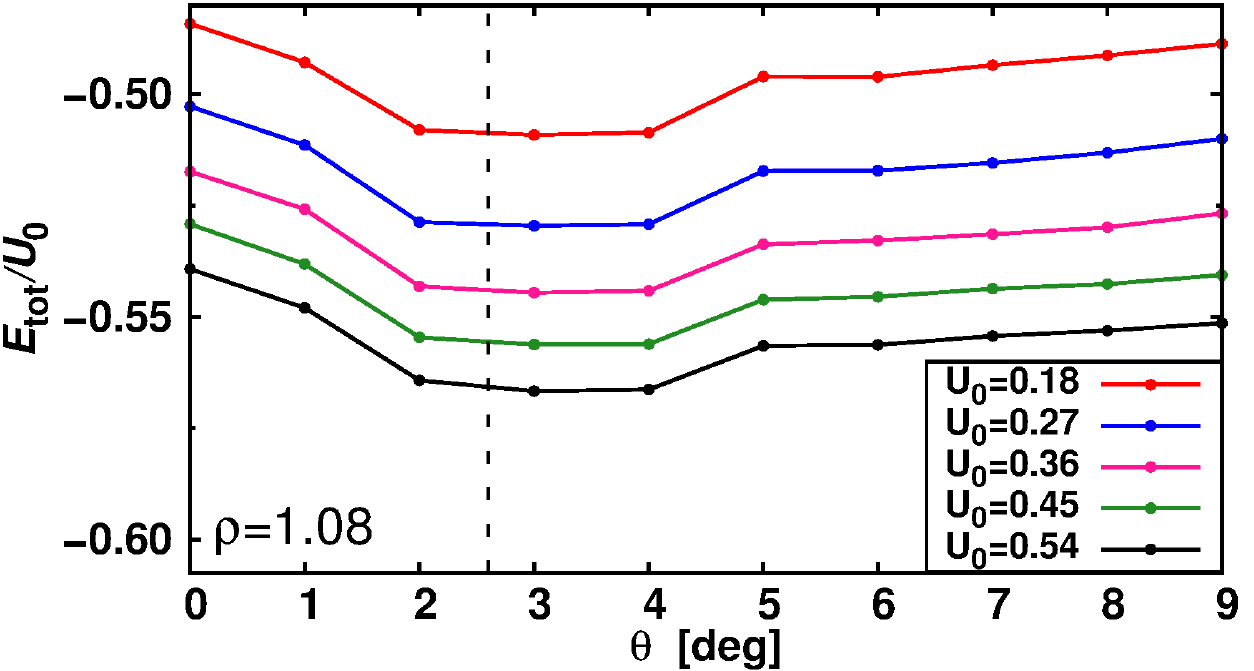}
 \caption{\label{fig3}
   Static structure optimization. Total energy per particle as a function
   of $\theta$ obtained in OBC at the overdense $\rho=1.08$, for different
   corrugations $U_0$.
   The dashed line indicates the expected value $\theta_{\rm NM}\simeq
   2.6^\circ$ from Eq.~(1) of the paper. }
 \end{center}
 \end{figure}
\section{Overdense incommensurate}
\begin{figure}[!t]
 \begin{center}
 \includegraphics[angle=0,width=0.45\textwidth]{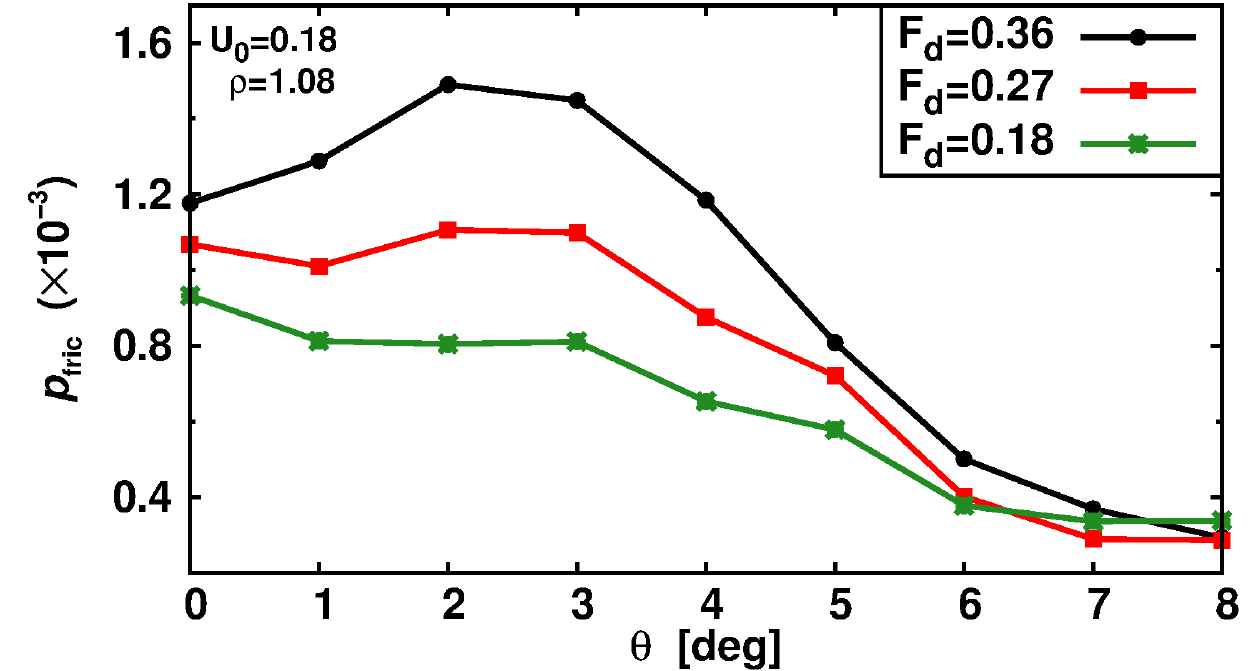}
 \caption{\label{fig4}
   Sliding dynamics in OBC.
   Dissipated friction power $p_{\rm fric}$ as a function of $\theta$
   obtained at $\rho=1.08$, external forces $F_d=0.18,\ 0.27,\ 0.36$, and
   corrugation $U_0=0.18$ ($T=0$). }
 \end{center}
\end{figure}

Figure~\ref{fig3} displays the total energy as a function of $\theta$ for
the overdense $\rho=1.08$. The theoretical NM angle is $\theta_{\rm NM}=2.6^\circ$.
The minimum is found in correspondence of $\theta_{\rm opt}\simeq 3^\circ$,
closer to the expected value than what was found in the UI regimes.
In this overdense case, moving toward the periphery the average colloid
spacing tends to get commensurate to the corrugation potential, favouring
smaller epitaxial angles.
In the OI regime the misfit stress is confined in local compressions rather
than dilations. Due to the exponential form of the repulsive interaction the cost of
compressions is higher, so that, at equal corrugation $U_0$, distortions
are weaker than in the UI case.
Consequently the dissipative sliding of the aligned and optimally
tilted configuration is not very different and the difference
in $p_{\rm fric}$ is reduced compared to the underdense regime.
Nonetheless, far away from $\theta_{\rm opt}$ a significant drop of
the dynamic friction is still observed, as shown in Fig.~\ref{fig4}.
\section{Temperature effects}
The NM theory \cite{NOV} assumes zero temperature.
Without attempting a rigorous investigation of the problem here we
anticipate, based on Eq.~(1) of the paper and on finite-temperature
simulations, that thermal effects do not affect the main conclusions
obtained at $T=0$.

Within a mean-field theory, such as a quasi-harmonic approximation, one may
still use $T=0$ results with thermally renormalized constants.
The two parameters defining $\theta_{\rm NM}$ are the mismatch ratio
$\rho=a_l/a_c$ and the sound velocity ratio $\gamma=c_L/c_T$.
The mismatch ratio is under control as one can always set the desired
incommensuration by fixing the average lattice spacing $\langle a_c\rangle$.
Of course $a_c$ will be characterized by a broader distribution at $T\ne 0$.
Changes in the sound velocities are also irrelevant if the effects of temperature on the crystal properties
act simply as a renormalization of the 2-body interaction.
Under this hypothesis $\gamma$ is expected to be weakly affected since, as
it turns out, its value is mainly determined by the symmetries of the
lattice \cite{note1}.
Therefore in a mean-field treatment, the rotated epitaxy is expected to
survive with only quantitative changes related to lattice thermal expansion.

\begin{figure}[!h]
 \begin{center}
 \includegraphics[angle=0,width=0.45\textwidth]{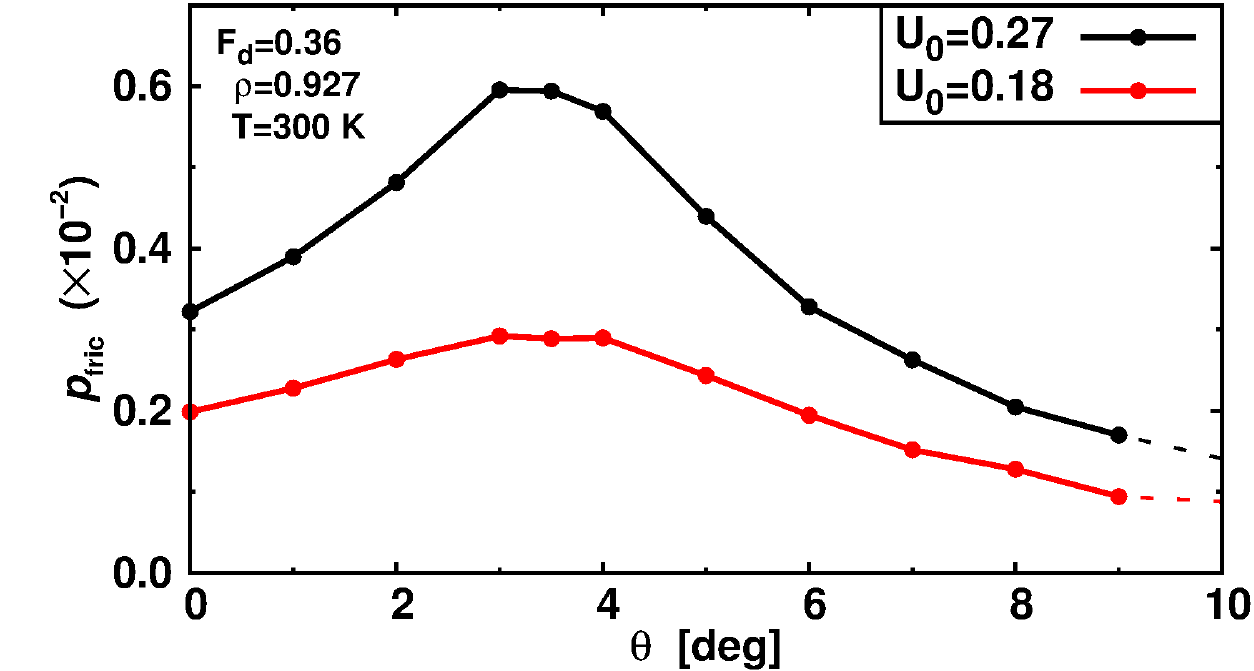}
 \caption{\label{fig5}
   The dissipated friction power $p_{\rm fric}$ as a function of $\theta$
   at $T_{\rm room}$. Simulations have been performed in OBC, fixing $\rho=0.927$, external
   force $F_d=0.36$ and corrugations $U_0=0.18,\ 0.27$.  }
 \end{center}
\end{figure}

The peak in the dynamic friction is connected with the increased
interdigitation of the monolayer within the substrate occurring at $\theta_{\rm opt}$.
At the optimal misalignment $\theta_{\rm opt}$, solitons are enhanced,
which is reflected in the larger distortions observed in the colloidal
crystal (Figs.~4 and 6b of the paper), and their motion is more dissipative.
At room temperature ($k_B T_{\rm room} \simeq 0.04$ in simulation units)
we observed that such solitonic superstructures are preserved for corrugations
$U_0>0.1$ and dynamic friction does indeed increase near the optimal
misalignement, as shown in Fig.~\ref{fig5}.
%
